# Approach to Toric Code Anyon Excitation, Indirect Effects of Kitaev Spin in Local Social Opinion Models

Yasuko Kawahata [†]

Faculty of Sociology, Department of Media Sociology, Rikkyo University, 3-34-1 Nishi-Ikebukuro,Toshima-ku, Tokyo, 171-8501, JAPAN.
ykawahata@rikkyo.ac.jp,kawahata.lab3@damp.tottori-u.ac.jp

**Abstract:** The study of Opinion Dynamics, which explores how individual opinions and beliefs evolve and how societal consensus is formed, has been examined across social science, physics, and mathematics. Historically based on statistical physics models like the Ising model, recent research integrates quantum information theory concepts, such as Graph States, Stabilizer States, and Toric Codes. These quantum approaches offer fresh perspectives for analyzing complex relationships and interactions in opinion formation, such as modeling local interactions, using topological features for error resistance, and applying quantum mechanics for deeper insights into opinion polarization and entanglement. However, these applications face challenges in complexity, interpretation, and empirical validation. Quantum concepts are abstract and not easily translated into social science contexts, and direct observation of social opinion processes differs significantly from quantum experiments, leading to a gap between theoretical models and real-world applicability. Despite its potential, the practical use of the Toric Code Hamiltonian in Opinion Dynamics requires further exploration and research.

**Keywords:** Opinion Dynamics, Quantum Information Theory, Toric Code Hamiltonian, Ising Model, Quantum Entanglement, Quantum Superposition, Theoretical Models in Social Dynamics, Kitaev Spin

## 1. Introduction

Opinion Dynamics, or the model of opinion formation, has been studied in various fields such as social science, physics, and mathematics. Traditional research on opinion dynamics is primarily based on models in statistical physics, particularly the Ising model. Recently, there have been attempts to apply concepts of quantum information theory to social sciences, including Graph States, Stabilizer States, and Toric Codes. The application of these quantum theoretical concepts to opinion dynamics could provide new perspectives.

### 1.1 History and Context of Opinion Dynamics

Opinion dynamics research aims to understand how individual opinions and beliefs change over time and how a common opinion or consensus is formed in society as a whole. Early models were inspired by physical systems like the Ising model, modeling opinions as spins and investigating how these spins interact and change opinions.

### 1.2 New Perspectives in Quantum Opinion Dynamics

Applying concepts of quantum information theory to opinion dynamics offers new perspectives:

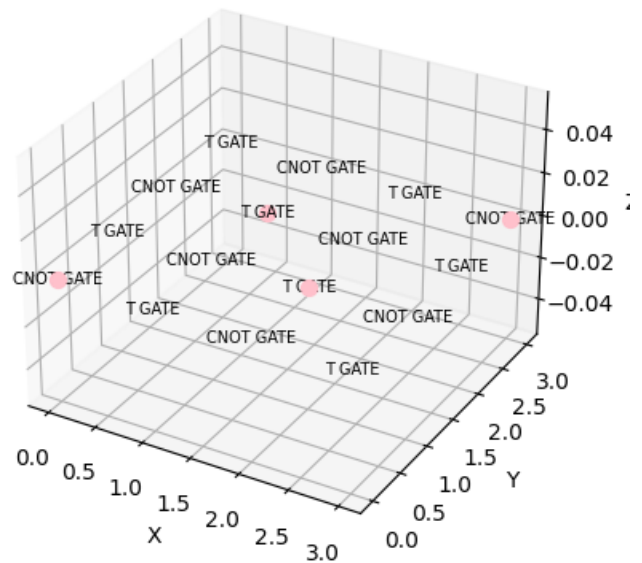

Fig. 1: Toric Code Torus (nxn) with Errors(pink)

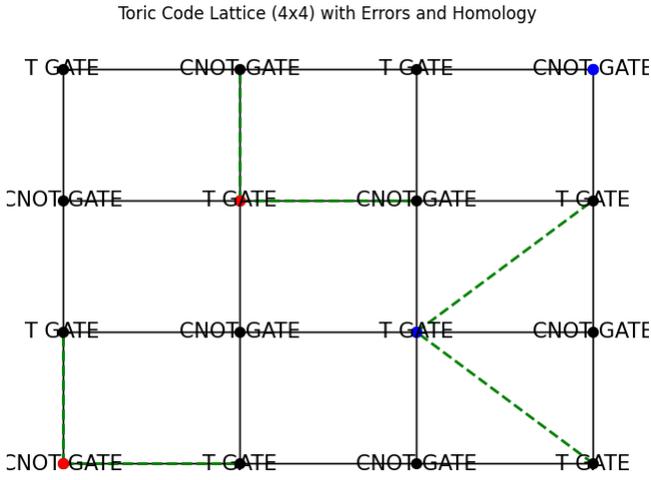

Fig. 2: Toric Code Lattice (nxn) with Errors, Homology

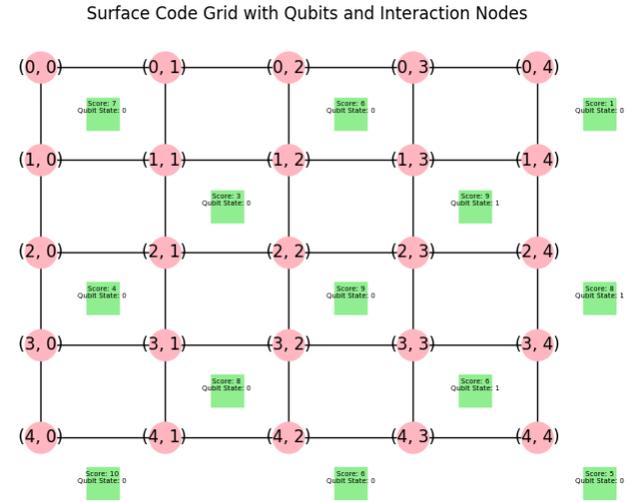

Fig. 3: Tile Graph of Interaction Coefficients:Quantum Toric Code Opinion Dynamics

(1) **Graph States and Stabilizer States:** These states can describe complex relationships between quantum bits. In opinion dynamics, this can be interpreted as the network of relationships and influences between people's opinions.

(2) **Topological Features of Toric Codes:** Toric codes exhibit resistance to errors. In the application of opinion dynamics, this can be interpreted as a mechanism to mitigate the impact of disagreements or misunderstandings on the opinion formation process.

(3) **Quantum-based Ising Model in Opinion Dynamics:** The Ising model, where spins take either up or down states (e.g., agree or disagree), can be enhanced in its quantum version to explore more complex dynamics of opinions, incorporating concepts of quantum superposition and entanglement.

## 1.3 Challenges in Application

However, there are challenges in applying quantum theory to social science problems:

**Complexity and Interpretation:** Quantum theory concepts are highly abstract, and their interpretation in the context of social science is often not intuitive.

**Empirical Validation:** Empirically validating models of quantum opinion dynamics is challenging, as the process of social opinion formation differs from quantum mechanical experiments and is difficult to observe directly.

**Gap between Theory and Practice:** There is often a significant gap between theoretical models and real-world social processes.

Quantum opinion dynamics has the potential to provide new perspectives in understanding opinion formation in social science. However, its application remains largely theoretical, and direct application to real social phenomena is still in its early stages. Future research will be important in linking these theories with empirical studies in social science.

To explain in detail the applications of Graph States, Stabilizer States, and Toric Codes in a quantum-based Ising model for opinion dynamics, it is necessary to explore how these quantum theoretical concepts can be applied to opinion dynamics.

### 1.3.1 Graph States in Opinion Dynamics

Graph States, where multiple quantum bits are entangled in a specific pattern, can be used in opinion dynamics to represent the process by which individual opinions (or beliefs) interact and influence the formation of collective opinions.

**Formation of Opinions:** Each quantum bit represents an individual opinion, and their interactions are represented by the graph state.

**Collective Influence:** Quantum entanglement is suitable for representing collective correlations and influences between opinions.

### 1.3.2 Stabilizer States in Opinion Dynamics

Stabilizer States, which remain invariant under a set of specific quantum gates, can be used to model the stability and consistency of opinions in opinion dynamics.

**Stability of Opinions:** Stabilizer states can represent groups of opinions that are stable under certain information or belief systems.

**Consistency and Shared Beliefs:** The properties of stabilizer states can help illustrate how shared beliefs and values maintain consistency within a group.

## 2. Toric Codes in Opinion Dynamics

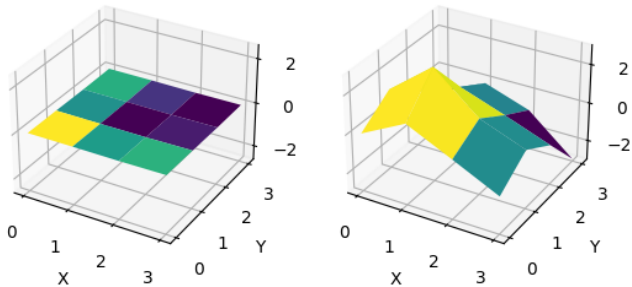

Fig. 4: Parameters of the local Kitaev model

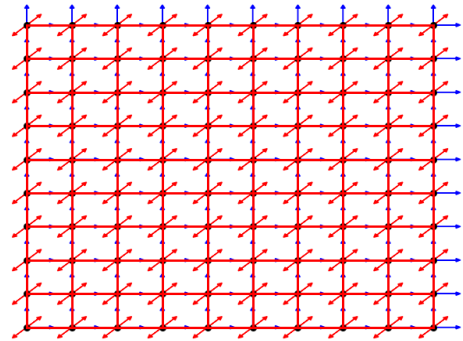

Fig. 5: Toric Code on a x Lattice: Star Operator, Red: Plaquette Operator

Toric Codes, which have topological quantum error-correcting properties, can be applied to model mechanisms that mitigate the impact of changes in opinions and misunderstandings in the opinion formation process.

**Change in Opinions:** The error-correcting properties of toric codes can demonstrate the resilience of social systems to changes in opinions and misunderstandings.

**Non-local Interactions in Social Networks:** The topological properties of toric codes are suitable for modeling the propagation and non-local interactions of opinions within social networks.

## Quantum-based Ising Model in Opinion Dynamics

Applying a quantum version of the Ising model to opinion dynamics can result in a model with the following characteristics:

**Polarization of Opinions:** In the quantum version of the Ising model, opinions can not only polarize but also take more complex states due to quantum superposition.

**Quantum Entanglement and Correlation of Opinions:** Quantum entanglement can be useful in expressing deep correlations and common consciousness within a group.

article [utf8]inputenc

## Application of the Toric Code Hamiltonian in Opinion Dynamics

The application of the Toric Code Hamiltonian to Opinion Dynamics, a model of social opinion formation, is a theoretically intriguing endeavor. The Toric Code Hamiltonian, fundamentally defined within the framework of quantum mechanics, describes interactions in a spin system, and its application as a metaphor in Opinion Dynamics is feasible.

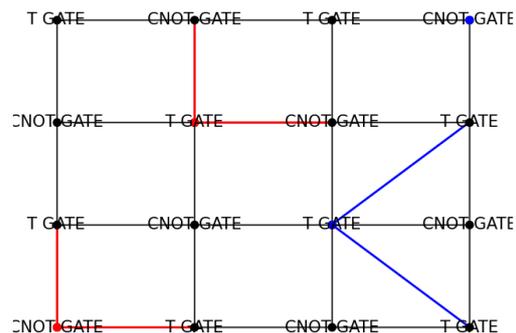

Fig. 6: Toric Code Lattice (nxn) with Errors

## The Hamiltonian of the Toric Code

The Hamiltonian of the Toric Code is expressed as follows:

$$H = -J_e \sum_s A_s - J_m \sum_p B_p$$

where,

$J_e$ and $J_m$ represent the strength of interactions.

$A_s$ is the product of Pauli $X$ operators for the four edges connected to the vertex $s$ (Star operator).

$B_p$ is the product of Pauli $Z$ operators around the face $p$ (Plaquette operator).

## Application to Opinion Dynamics

The significance of using the Toric Code Hamiltonian in Opinion Dynamics can be considered as follows:

(1) **Modeling Local Interactions:**

**Star Operator** $A_s$: Can be used to model how each individual (or opinion) interacts with their neighbors. This can represent the influence and harmony of opinions within a social network.

**Plaquette Operator** $B_p$: Can model the consistency and commonality of opinions within a social group or community.

(2) **Energy Minimization and Stability of Opinions:**

The ground state of the Toric Code model represents a state of minimized energy. In Opinion Dynamics, this could correspond to the stability of opinions or a state of shared beliefs.

(3) **Utilizing Topological Properties:**

The topological properties of the Toric Code can be used to understand more complex dynamic characteristics in opinion formation and change, as well as non-local interactions within social networks.

## Conclusion and Challenges

Such applications have the potential to offer new methods for understanding the dynamics of social opinion formation, but they also present several challenges:

**Complexity:** The Toric Code Hamiltonian is defined within the framework of quantum mechanics, and its direct application to social sciences is complex and not intuitive.

**Difficulty in Interpretation:** Applying concepts of quantum mechanics in the context of social science involves very challenging problems.

**Lack of Empirical Validation:** It is difficult to empirically validate the effectiveness and appropriateness of such models.

In conclusion, while the application of the Toric Code Hamiltonian to Opinion Dynamics is theoretically interesting, its feasibility and utility require further research.

## 2.1 Topological Features in Opinion Dynamics

Topological Features in Opinion Dynamics Because of their very interesting topological properties, Kitaev spins have been applied in different areas of physics, such as in opion dynamics. In general, Kitaev spins are often used to study quantum topological materials, but the concept of topological properties can also be applied to different physical systems.

Kitaev spin refers to one particular model or Hamiltonian (energy representation) of quantum spin systems proposed by Alexei Kitaev. This model plays a very important role in modern quantum physics and computer science, including the study of topological quantum computers and topological quantum matter.

Kitaev spin is one of the models describing quantum spin systems. Quantum spin is one of the fundamental degrees of freedom in quantum mechanics and can have values such as spin 1/2 or spin 1. It usually describes the interaction of spins with two values of spin (spin 1/2). It does not have the usual interaction terms and has a very specific interaction pattern. In this model, neighboring spins do not interact directly with each other and there is a nonlocal interaction. Also, quantum phase transitions are said to occur. In addition, topological properties are of interest. In this model, topological properties appear in the energy spectrum, such as topological chirality and nontrivial edge modes. This could be applied as a basis for topological matter and topological quantum computers. 4. Topological chirality is a topologically protected edge state in some physical systems, with a characteristic gap in the energy spectrum. This is a useful property in information protection and quantum computation. It has a wide range of applications in topological quantum computers such as the toric code, topological insulators, and topological materials. In particular, it plays an important role in the field of quantum computation in studies related to error correction and information protection.

Opinion dynamics is applied in areas such as social networks and information diffusion, where properties of information propagation and interaction are studied. Applying the Kitaev model to opinion dynamics may be beneficial from the following perspectives

It is important in terms of topological protection. Kitaev models are known to have topological phases and have interesting properties for edge states. In the context of opinion dynamics, there may be topological properties of information propagation and protected states that may provide new insights into information propagation and diffusion. In the Kitaev model, special exotic particles called mejanafermions appear. These Mejana fermions have topological chirality and may have nontrivial effects on information propagation and interactions. 3.

In terms of quantum phase transitions the Kitaev model may be used to study topological quantum phase transitions. Analogies to phase transitions may also emerge in opinion dynamics and may help to understand phase transitions and critical phenomena in information diffusion.

However, the application of the Kitaev model to opinion dynamics may require appropriate modifications and extensions regarding the specific conditions of the model, interaction patterns, and network structure. In addition, it is important to collaborate with researchers in topological properties and opinion dynamics to understand the properties of the Kitaev model and apply it to the context of opinion dynamics.

In short, the application of the topological properties of the Kitaev model to opinion dynamics is a promising research direction because it has the potential to provide new insights and understanding, but the specific approach will depend on the research objectives and problems, and this paper will be limited to exploring the possibilities.

# 3. Opinion Dynamics adopted Toric Node, Previews Reference

## 3.1 Topological quantum computing

The following is a list of research papers on topological quantum computing.

Kitaev, A. Y. (2003) introduces fault-tolerant quantum computation using particles called anyons, which possess non-trivial statistics.

Nayak, C., Simon, S. H., Stern, A., Freedman, M., and Das Sarma, S. (2008) focuses on non-Abelian anyons, topological particles, and discusses the potential of topological quantum computation.

Dennis, E., Kitaev, A., Landahl, A., and Preskill, J. (2002) discusses the realization of reliable quantum memory using topological error correction codes.

Sarma, S. D., Freedman, M., and Nayak, C. (2015) centers on Majorana zero modes, a type of topological excitation state, and its application in topological quantum computation.

These papers represent significant contributions to the foundational concepts and theories in topological quantum computing. Topological quantum computing holds innovative potential in terms of improving the reliability of quantum bits and error correction. These studies contribute to its development." These papers focus on important concepts and their realization in the field of topological quantum computing. In Kitaev's paper, the possibility of fault-tolerant quantum computation using anyons is discussed. Dennis and his colleagues elaborate on how to construct reliable quantum memory using topological error correction codes. Nayak et al. comprehensively explain the connection between non-Abelian anyons and topological quantum computation, while Sarma's paper explores the potential of topological quantum computation utilizing Majorana zero modes.

These papers provide a theoretical foundation for topological quantum computing and represent significant steps towards the development of new quantum computing architectures. Topological quantum computing offers an innovative approach to quantum bit error correction and information storage, and these studies contribute to its advancement.

And Kitaev's paper, the focus is on fault-tolerant quantum computation using anyons. This emphasizes the importance of error correction in quantum computing and proposes a method to achieve reliable computations using anyons.

Bravyi and König's research focuses on the classification of topologically protected gates for local stabilizer codes. This deepens our understanding of fundamental gate operations in topological quantum computers and provides valuable information for their implementation.

In Barkeshli and Klich's paper, they investigate the fractionalization of Majorana fermions with non-Abelian statistics at the edges of abelian quantum Hall states. This expands the possibilities of topological quantum computing by utilizing particles with different statistical behaviors.

Finally, Bonderson, Kitaev, and Shtengel propose a new topological quantum error correction code called Cat-Code Quantum Codes. This offers an innovative approach to error correction and contributes to improving the reliability of topological quantum computing.

These papers provide crucial insights into the fundamental theory and implementation of topological quantum computing, contributing to the future development of quantum computing technology.

## 3.2 Quantum memory and quantum error correction

And The following papers provide an overview of research on topological quantum memory and quantum error correction. Dennis et al.'s research focuses on "topological quantum memory," with a theoretical approach centered on the "Toric Code," one of the topological error correction codes. The Toric Code serves as a crucial foundation for error detection and correction in physical implementations.

In Alicki and Fannes' book, "Quantum dynamical systems," the emphasis is on dynamical processes in quantum mechanics. Understanding the time evolution and dynamics of quantum systems is essential for quantum error correction and quantum information processing.

Wang, Fowler, and Hollenberg's research explores quantum computing using the "Surface Code" with error rates exceeding 1%. The Surface Code is one of the topological codes that enables efficient error correction.

Lastly, Poulin and Chung's study delves into research on "Toric Code" and "Topological Color Codes" in three-dimensional space. These topological codes are relevant to error correction and topological quantum computation in three-dimensional space.

The following paper list provides references for textbooks on nuclear physics.

The 1989 textbook by Brink and Satchler, titled "Angular Momentum," offers a detailed explanation and computational methods for angular momentum in nuclear physics.

The 1975 textbook by Bohr and Mottelson, titled "Nuclear Structure," provides comprehensive information about the structure and properties of nuclei in nuclear physics.

The 1980 textbook by Ring and Schuck, titled "The Nuclear Many-Body Problem," focuses on the many-body problem in nuclear physics.

The 2003 textbook by Fetter and Walecka, titled "Quantum Theory of Many-Particle Systems," explains fundamental quantum theory related to many-particle systems.

The 1986 textbook by Blaizot and Ripka, titled "Quantum Theory of Finite Systems," offers theory and computational methods for finite quantum systems.

These textbooks focus on important topics in nuclear physics and serve as valuable resources for learners and researchers interested in the fundamental theory and practice of nuclear physics. They are valuable literature for those interested in the basic principles of nuclear physics.

And these papers offer crucial insights into the theory and implementation of topological quantum memory and error correction, contributing to the advancement of future quantum computing technologies.

### 3.3 Quantum memory and quantum error correction2

In Shor's 1995 paper, a scheme was proposed to reduce decoherence (information loss) in the memory of quantum computers.

Steane's 1996 paper, titled "Error Correction Codes in Quantum Theory," provides the fundamental theory of error correction codes in quantum computing.

In the 1997 paper by Knill, Laflamme, and Zurek, they discuss "Resilient quantum computation," focusing on error correction and improving reliability.

Calderbank, Rains, Shor, and Sloane's 1997 paper is about "Quantum error correction and orthogonal geometry," explaining the geometric approach to quantum error correction codes.

Preskill's 1998 paper discusses "Reliable quantum computers" and explores approaches to achieving reliable quantum computers.

Finally, Knill and Laflamme's 1998 paper focuses on "A theory of quantum error-correcting codes," providing a theoretical foundation for quantum error correction codes.

These papers emphasize the importance of error correction in quantum computing and contribute to laying the groundwork for the development of reliable quantum computers.

### 3.4 Quantum field theory

The following list of papers provides an explanation of quantum field theory. This list includes important research papers and textbooks related to the theory of quantum fields.

Abrikosov, Dzyaloshinski, and Lifshitz authored a book in 1975 titled "Methods of Quantum Field Theory in Statistical Physics," which comprehensively explains the methodology of quantum field theory in statistical physics.

In 2000, Mahan authored the book "Many-Particle Physics," covering fundamental concepts and methods in many-body physics.

The 2003 book by Fetter and Walecka, "Quantum Theory of Many-Particle Systems," focuses on quantum theory of many-body systems and addresses topics relevant to applications in solid-state physics.

Rammer's 2007 book, "Quantum Field Theory of Non-equilibrium States," centers on non-equilibrium quantum field theory, elucidating methods for describing non-equilibrium states.

Altland and Simons authored the book "Condensed Matter Field Theory" in 2010, concentrating on the application of quantum field theory in condensed matter physics and covering topics like topological matter.

Schwinger's 1961 paper, "Brownian motion of a quantum oscillator," conducts theoretical research on the Brownian motion of quantum oscillators.

Kadanoff and Baym's 1962 book, "Quantum Statistical Mechanics: Green's Function Methods in Equilibrium and Nonequilibrium Problems," provides a detailed explanation of Green's function methods for equilibrium and nonequilibrium problems in quantum statistical mechanics.

These references serve as valuable resources for researchers and students interested in quantum field theory and many-body physics, contributing to a deeper understanding of this field.

### 3.5 Quantum entanglement

The following list of papers provides valuable insights into the research on quantum entanglement.

The 2008 paper by Amico, Fazio, Osterloh, and Vedral, titled "Entanglement in many-body systems," focuses on quantum entanglement in many-body systems, comprehensively explaining its properties and theoretical approaches.

In 2004, Calabrese and Cardy discussed the relationship between quantum field theory and entanglement entropy in their paper "Entanglement entropy and quantum field theory."

The 2009 paper by the Horodecki brothers, "Quantum entanglement," provides a comprehensive overview of the fundamental properties, classification, and applications of quantum entanglement.

In their 2010 paper, "Area laws for the entanglement entropy," Eisert, Cramer, and Plenio delve into the area laws governing entanglement entropy, offering detailed explanations of the properties of entanglement in quantum many-body systems.

Calabrese and Cardy's 2009 paper, "Entanglement entropy and conformal field theory," discusses the relationship between entanglement entropy and conformal field theory.

These papers serve as valuable references for physicists and researchers interested in the theory and experimental study of quantum entanglement, providing a foundation for quantum information science and the study of many-body systems.

### 3.6 Jordan-Wigner transformation

The following list of papers provides an explanation of research on the Jordan-Wigner transformation:

The 1928 paper by Jordan and Wigner, titled "Über das Paulische Äquivalenzverbot," conducted research on Pauli's exclusion principle. This was a crucial discovery concerning the exchange rules of fermions in quantum mechanics and laid the foundation for the Jordan-Wigner transformation.

In 2002, the paper by Bravyi and Kitaev, titled "Fermionic quantum computation," delved into research on fermionic quantum computation. The Jordan-Wigner transformation serves as an essential tool for handling Hamiltonians of fermionic systems in quantum computation.

In 2014, Larsson's paper, titled "A Short Review on Jordan-Wigner Transforms," provided a concise review of the Jordan-Wigner transformation.

In 2006, Mazziotti's paper, titled "Jordan-Wigner transformation with arbitrary locality," discussed methods for extending the Jordan-Wigner transformation to cases with arbitrary locality.

These papers collectively offer valuable insights into the Jordan-Wigner transformation and its applications, contributing to research in fields such as quantum computation and quantum chemistry.

### 3.7 Overview of Majorana Fermions

These papers introduce research on Majorana fermions in solid-state systems. In Alicea's 2012 paper, a new direction for exploring Majorana fermions within solid materials is proposed, offering fresh perspectives on the study of these intriguing particles.

Kitaev's 2001 paper, on the other hand, puts forth theoretical foundations for non-Abelian Majorana fermions within quantum wires, marking a significant milestone in Majorana fermion theory.

In the 2008 paper by Aguado and Vidal, discussions revolve around entanglement renormalization and Majorana fermions, hinting at the connection between entanglement and topological properties.

Finally, the 2010 paper by Lutchyn, Sau, and Das Sarma focuses on Majorana fermions in semiconductor-superconductor heterostructures and topological phase transitions. These studies suggest the potential significance of Majorana fermions in quantum computing and topological quantum information processing.

These papers collectively provide crucial insights into the exploration of new physical phenomena and offer significant prospects in the field of quantum information processing, highlighting the pivotal role that Majorana fermion research is poised to play in future scientific endeavors.

### 3.8 Green's function research in the fields of many-body physics

These papers provide an overview of classical research in the fields of many-body physics and statistical physics, representing significant achievements that laid the foundations of these disciplines. Abrikosov's work serves as a comprehensive guide introducing methods and techniques in quantum field theory for statistical physics, making it a valuable resource for researchers in the field.

Mahan's book focuses on the theoretical and experimental applications of many-body effects in solid-state physics, exploring fundamental concepts in many-body physics. Fetter and Walecka's publication provides a detailed explanation of quantum mechanics for many-body systems, offering the fundamentals of many-body physics.

Additionally, Rammer's book delves into quantum field theory in non-equilibrium states, addressing time-dependent issues in statistical physics. Altland and Simons comprehensively explain the field theory of condensed matter systems, tackling advanced topics in statistical physics.

The papers by Schwinger and Kadanoff showcase classical approaches in statistical physics and quantum field theory, playing crucial roles in establishing the foundations of many-body physics and statistical physics.

These sources are valuable references for researchers and students in many-body physics and statistical physics, and they hold significant value for anyone interested in classical research in these fields.

### 3.9 Band structure of excitation energy

The following papers provide an overview of research on the band structure of excitation energy in solid-state physics and electronic structure calculations. Excitation energy is a crucial indicator for understanding the properties of materials associated with electronic excitations and plays a significant role in material design and property prediction.

In Kohn-Sham Density Functional Theory (DFT), a commonly used method in electronic structure calculations, the paper by Perdew-Zunger extensively investigates the band structure of excitation energy within DFT and proposes improvements to approximation methods.

Hedin's paper introduces advanced calculation methods that consider strong electronic correlation effects, known as the GW approximation, contributing to the enhancement of accuracy in the band structure of excitation energy.

The research by Hybertsen and Louie suggests a method combining the GW approximation and the Bethe-Salpeter equation (BSE) as an approach for calculating excitation energy, contributing to the understanding of electronic excitation properties in materials.

These papers demonstrate significant advancements in the calculation methods for excitation energy and the understanding of electronic excitation properties in materials. They have a substantial impact on fundamental research in materials science and solid-state physics.

## 4. Discussion

### 4.1 Introduce of The Toric code

The Toric code is a model that plays an important role in quantum error correction and topological quantum computation. The Hamiltonian of this model describes a spin system on a lattice and includes specific local interactions. Generally, a two-dimensional square lattice is considered, with spin-1/2 particles (qubits) placed on each edge.

The Hamiltonian of the Toric code is defined as follows:

$$H = -J_e \sum_s A_s - J_m \sum_p B_p$$

where,

$J_e$ and $J_m$ are constants representing the strength of the interactions.

The sums are taken over all vertices $s$ and plaquettes $p$ (faces of the lattice) on the lattice.

$A_s$, known as the 'star operator', acts on the spins on the four edges connected to vertex $s$. Specifically, $A_s = \prod_{i \in star(s)} \sigma_i^x$, where $\sigma_i^x$ is the Pauli X matrix acting on the $i$th spin.

$B_p$, referred to as the 'plaquette operator', acts on the spins around the boundary of the plaquette $p$. This is defined as $B_p = \prod_{i \in boundary(p)} \sigma_i^z$, where $\sigma_i^z$ is the Pauli Z matrix acting on the $i$th spin.

A characteristic feature of this model is that its ground state is topologically protected, making it highly robust against local perturbations. Furthermore, the excitations exhibit very peculiar behaviors, and it is known that these excitations play a significant role in topological quantum computation.

The Toric Code is a type of topological quantum error-correcting code defined on a lattice on a two-dimensional torus. To understand the Toric Code, it is important to consider the non-commutativity of the Pauli matrices $X$ and $Z$, and their actions along specific closed loops on the torus.

### 4.2 Fundamental Elements of the Toric Code
### 1. Lattice Structure

The Toric Code is based on a two-dimensional square lattice arranged in a torus (donut shape). Quantum bits (qubits) are placed on each edge.

### 2. Action of Pauli Operators

The $X$ operator (bit flip) and $Z$ operator (phase flip) act on each edge. In the Toric Code, the non-commutative relationship of these operators is crucial. That is, $X$ and $Z$ are non-commutative ($XZ = -ZX$) only when they act on the same edge, but commutative when acting on different edges.

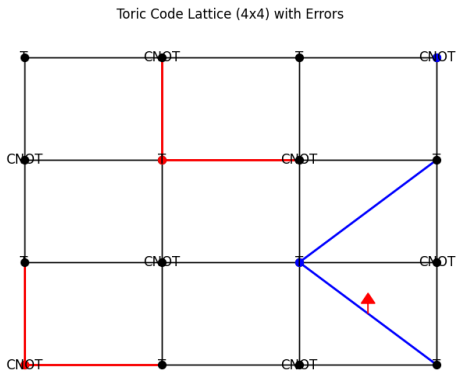

Fig. 7: Toric Code on a x Lattice: Star Operator, Red: Plaquette Operator

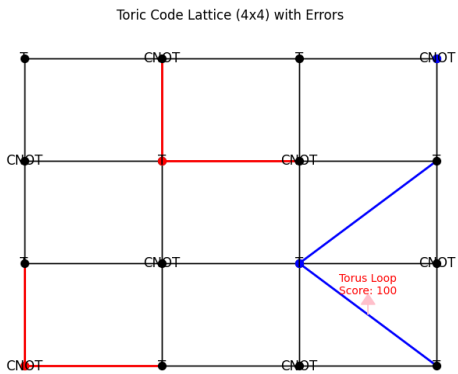

Fig. 8: Toric Code on a x Lattice: Star Operator, Red: Plaquette Operator

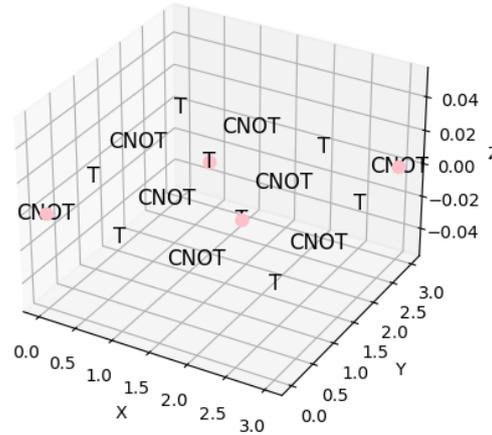

Fig. 9: Toric Code Lattice (nxn) with Errors, Homology

## 3. Star and Plaquette Operators

The star operator $A_s$ and the plaquette operator $B_p$ are associated with the vertices and faces of the lattice, respectively. $A_s$ applies the $X$ operator to all edges connected to a vertex, and $B_p$ applies the $Z$ operator to all edges forming the boundary of a face.

### Basis Vectors of the Toric Code

The basis vectors of the Toric Code are generated by specific loop operators. Here, $L(X)$ represents the product of $X$ operators along a loop that encircles one direction of the torus, and $M(X)$ represents the product of $X$ operators along a loop in the other direction of the torus. Similarly, $L(Z)$ and $M(Z)$ correspond to $Z$ operators.

The ground state $\Omega$ is stable (i.e., unchanged) by all $A_s$ and $B_p$ operators. The four basis vectors of the Toric Code are:

1. $\Omega$ **(state with all qubits in 0)**

2. $L(X)\Omega$

3. $M(X)\Omega$

4. $L(X)M(X)\Omega$

These states represent different topological sectors. Since $L(X)$ and $M(X)$ are loops along different directions of the torus, these operators are commutative with each other, generating independent topological sectors. These basis vectors are orthogonal to each other, demonstrating the topological properties of the Toric Code.

## Outline of the Proof

The proof that this basis forms a complete basis for the Toric Code is done by showing that these states are invariant under all $A_s$ and $B_p$ operators, and are orthogonal to each other. Furthermore, it is important that these states reflect the topological properties of the torus. This is understood from the fact that different loop operators defined on the torus correspond to different topological sectors.

In the Toric Code, anyons appear as excitations. These excitations arise by changing the eigenvalues of the star operator $A_s$ and the plaquette operator $B_p$. There are mainly two types of anyons in the Toric Code: 'electric charge' (e-type) anyons and 'magnetic flux' (m-type) anyons.

### 4.3 Electric Charge (e-type) Anyons

Electric charge anyons are created by changing the eigenvalues of the star operator $A_s$. The star operator is the product of Pauli $X$ operators for the four edges connected to a vertex. Normally, the eigenvalue of $A_s$ is +1, but when a Pauli $Z$ operator acts on an edge, the eigenvalues of the two star operators connected to that edge change to $-1$. This creates a pair of 'electric charge' anyons.

## Magnetic Flux (m-type) Anyons

Magnetic flux anyons are created by changing the eigenvalues of the plaquette operator $B_p$. The plaquette operator is the product of Pauli $Z$ operators for the edges forming the boundary of a face. Normally, the eigenvalue of $B_p$ is +1, but when a Pauli $X$ operator acts on an edge, the eigenvalues of the two plaquette operators connected to that edge change to $-1$. This creates a pair of 'magnetic flux' anyons.

## Non-Abelian Statistics of Anyons

The anyons in the Toric Code exhibit non-Abelian statistics. This means that when one type of anyon moves around another type, the overall quantum state changes. This is one of the reasons why the Toric Code plays an important role in topological quantum error correction and quantum computation.

## Graphical Explanation

Imagine a two-dimensional square lattice with quantum bits placed on each edge. Electric charge anyons appear at a vertex when a $Z$ operator acts on one or more of the four edges connected to that vertex. Magnetic flux anyons appear on a face when an $X$ operator acts on one or more of the edges surrounding that face.

Anyons are always created in pairs, and when one moves to a different location, a new operator action is required along its path. This means that anyons have specific topological

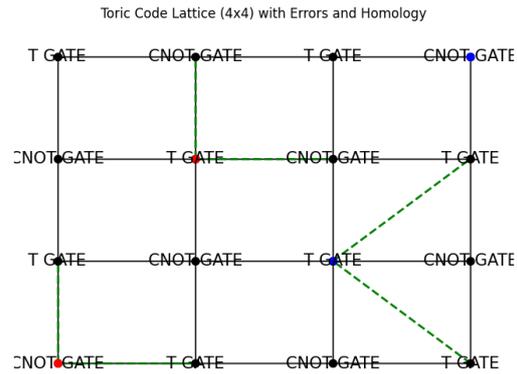

Fig. 10: Toric Code Lattice (nxn) with Errors, Homology

properties, and their exchange statistics differ from those of conventional bosons and fermions.

In the Toric Code, the process of 'filling holes', i.e., correcting defects (errors), is crucial as part of quantum error correction. This process involves the use of CNOT gates, T gates, and homology theory (a field of mathematics used to analyze topological properties). Here, we briefly explain how these concepts are involved in error correction for the Toric Code.

## Stabilizers and Defects

In the Toric Code, stabilizers (the star operators $A_s$ and the plaquette operators $B_p$) define the system's ground state. When a stabilizer's eigenvalue becomes $-1$, it indicates a defect (error). This defect is like a 'hole' in the Toric Code.

## Use of CNOT Gates

The CNOT gate (Controlled NOT gate) is used to create entanglement between qubits. In the Toric Code, CNOT gates play a crucial role in the error correction process. In particular, they are used to detect the location of errors and transfer that information to auxiliary qubits.

## Use of $T$ Gates

The T gate (/8 gate) is a gate that slightly changes the phase of a qubit, enabling more complex quantum algorithms. The specific use of the T gate in the Toric Code depends on the context of error correction but is generally used for fine-tuning quantum states.

## Homology and Error Correction

Homology theory plays a central role in error correction in the Toric Code. Errors can be interpreted as loops on the torus, and these loops are classified by homology groups. To correct errors, it is necessary to 'close' these loops appropriately. This is a computation done considering topological properties.

**Computational Process**

The specific computational process starts with assessing the current state of the system and identifying where defects (errors) exist. This is usually achieved by measuring stabilizers using auxiliary qubits. Then, based on the information obtained, errors are corrected using CNOT and T gates. This process is conducted while considering topological information using homology theory.

This process is a general approach in the field of quantum error correction and is particularly important for topological codes like the Toric Code. However, actual implementation is highly complex and greatly depends on both the hardware and software characteristics of quantum computers.

# 5. Toric Code Anyon Excitation and Opinion Dynamics Modeling

In this section, we propose the basic equations and parameters for describing the excitation of anyons in the toric code and modeling opinion dynamics using these excitations. However, the specific equations and parameters required may vary depending on the problem setting and system, so the following is a general approach. Customization will be necessary to match specific models.

## 5.1 Toric Code Anyon Excitation

In the toric code, the excitation of anyons is related to the eigenvalues of the star operator and the plaquette operator. The basic equation for describing the excitation of anyons is as follows:

$$H = -J \sum_{\text{e-type}} (\sigma_z) - K \sum_{\text{m-type}} (\sigma_x) \quad (1)$$

Here, $H$ is the Hamiltonian (energy operator), $J$ is the coefficient of the star operator, and $K$ is the coefficient of the plaquette operator. $\sigma_z$ and $\sigma_x$ are the z-component and x-component of the Pauli matrices, respectively.

## 5.2 Opinion Dynamics Modeling

To model opinion dynamics, we need to describe the dynamics of anyons on the toric code. The dynamics of anyons are usually modeled using Monte Carlo methods or quantum Monte Carlo methods. Below is the equation for a general Monte Carlo method:

$$P(\text{state } s \to s') = \min\left[1, \exp\left(-\beta(E(s') - E(s))\right)\right] \quad (2)$$

Here, $P(\text{state } s \to s')$ is the transition probability from state $s$ to $s'$, $E(s)$ is the energy of state $s$, and $\beta$ is the inverse temperature parameter. This probability is used to simulate the dynamics of anyon excitations.

## 5.3 Parameters

The parameters required for the model include $J$, $K$, $\beta$, the number of simulation steps, and others. These parameters need to be adjusted to match the specific system or problem. Additionally, consideration must be given to initial conditions and boundary conditions.

## 5.4 Incorporating Syndrome Measurement into Model Calculations

To incorporate syndrome measurement into the parameters of model calculations, it is necessary to set the parameters based on the specific system and syndrome measurement algorithm of the toric code. Below is an outline of a general approach, but the specific implementation should be adjusted to match the system.

### 1. Introduction of Syndrome Measurement

In the toric code, syndrome measurements are performed to detect errors. Syndrome measurement involves performing parity measurements known as X-syndromes and Z-syndromes. Introduce quantum circuits or classical algorithms to calculate these syndromes.

### 2. Error Correction Based on Syndromes

Based on the results of syndrome measurements, error correction operations are performed. Choose a specific error correction algorithm (e.g., the bit-flip method, the majority-flip method, etc.) and identify and correct errors from the syndromes.

### 3. Incorporating Syndrome Measurement into Model Calculations

Syndrome measurements can be incorporated into the parameters of model calculations. Depending on the specific approach, consider the following steps:

### Introduce Error Operators into the Hamiltonian

Incorporate error correction operations based on syndrome measurements into the Hamiltonian. The error operators include terms corresponding to X-syndromes and Z-syndromes.

## Probability Calculations with Monte Carlo Methods

Based on the results of syndrome measurements, use Monte Carlo methods or similar techniques to calculate the probability distribution of the state after error correction.

## Parameter Adjustment

Incorporate parameters related to syndrome measurement (e.g., the success probability of syndrome measurement) into the parameters of model calculations and perform simulations.

### 5.5 Proposed Equations for Model Calculations Including Anyon Excitation and Syndrome Measurement in the Toric Code

#### 1. Equation for Anyon Excitation in the Toric Code

Anyon excitation in the toric code is related to the eigenvalues of the star operator and plaquette operator. Taking into account the conditions for e-type and m-type, we can express the Hamiltonian as follows:

$$H_1 = -J \sum_{\text{e-type}} (\sigma_{zi}\sigma_{zj}) - K \sum_{\text{m-type}} (\sigma_{xi}\sigma_{xj}) \quad (3)$$

Here, $H_1$ is the Hamiltonian representing anyon excitation, $J$ is the coefficient of the star operator, and $K$ is the coefficient of the plaquette operator. $\sigma_{zi}$ and $\sigma_{xi}$ are the z-component and x-component of the Pauli matrices, respectively. $\sum_{\text{e-type}}$ and $\sum_{\text{m-type}}$ represent sums over pairs of anyons that satisfy the corresponding conditions.

#### 2. Equation for Syndrome Measurement

Syndrome measurement consists of measurements of X-syndromes and Z-syndromes. To represent them, we introduce the following equations:

$$H_2 = -h_x \sum_{X\text{-syndrome}} -h_z \sum_{Z\text{-syndrome}} \quad (4)$$

Here, $H_2$ is the Hamiltonian representing syndrome measurement, and $h_x$ and $h_z$ are coefficients related to the measurements of X-syndromes and Z-syndromes, respectively. $\sum_{X\text{-syndrome}}$ and $\sum_{Z\text{-syndrome}}$ represent sums over the corresponding syndrome measurements.

#### 3. Hamiltonian for the Entire Model

The Hamiltonian for the entire model is expressed by combining the terms for anyon excitation and syndrome measurement as follows:

$$H = H_1 + H_2 \quad (5)$$

This constructs a model of the toric code that includes the interaction between anyon excitation and syndrome measurement.

By adjusting parameters such as $J$, $K$, $h_x$, and $h_z$ according to specific systems or problem settings, you can model the dynamics of syndrome measurement and anyon excitation. Furthermore, you can perform simulations using these Hamiltonians.

## 6. Modeling the Behavior of Specific Local Interactions with Opinion Dynamics

The toric code is a model that plays a crucial role in quantum error correction and topological quantum computation. Its Hamiltonian describes a lattice-based spin system. Below, we present the Hamiltonian of the toric code and propose a formula to replace the behavior of specific local interactions with opinion dynamics.

The Hamiltonian of the toric code is typically composed of terms involving the star operator and the plaquette operator. These operators are defined on the lattice of the toric code and represent specific interactions. The star operator is associated with edges around vertices, and the plaquette operator is associated with edges around faces. The Hamiltonian is expressed as follows:

$$H = -J \sum_{\text{s-type}} -K \sum_{\text{p-type}} \quad (6)$$

Here, $H$ is the Hamiltonian, $J$ is the coefficient of the star operator, and $K$ is the coefficient of the plaquette operator. $\sum_{\text{s-type}}$ represents the sum over star operators, and $\sum_{\text{p-type}}$ represents the sum over plaquette operators. The s-type and p-type indices are used to apply specific conditions to the star and plaquette operators, respectively.

To model the behavior of specific local interactions using opinion dynamics, you can consider a formula like the following:

$$H_{\text{opinion}} = -J_{\text{opinion}} \sum_{\text{s-type}} -K_{\text{opinion}} \sum_{\text{p-type}} \quad (7)$$

Here, $H_{\text{opinion}}$ is the new Hamiltonian replaced by opinion dynamics. $J_{\text{opinion}}$ and $K_{\text{opinion}}$ are parameters set in opinion dynamics, controlling the strength of local interactions.

By setting specific conditions for s-type and p-type, you can model the behavior of specific local interactions. For example, by adding conditions to s-type that consider the distance or direction between specific qubits, you can control the local interactions between spins.

Depending on the specific problem setting and system, you can adjust parameters such as $J_{\text{opinion}}$, $K_{\text{opinion}}$, s-type, and p-type to customize the toric code model using opinion dynamics.

## 6.1 Model Considering the Behavior of Anyons in the Toric Code

Model considering the behavior of anyons (charge e-type and flux m-type) in the toric code, we propose a Hamiltonian that includes charge (e-type) and flux (m-type) anyons. Anyons are associated with changes in the eigenvalues of star operators and plaquette operators, and different terms are considered to model the behavior of charge and flux anyons.

Proposed Hamiltonian:

$$H = -J_s \sum_{\text{s-type}} -J_p \sum_{\text{p-type}} + H_e + H_m \qquad (8)$$

Where:

- $J_s$ represents the strength of interactions associated with star operators. - $J_p$ represents the strength of interactions associated with plaquette operators. - $H_e$ is the term for charge (e-type) anyons, associated with changes in the eigenvalues of star and plaquette operators. - $H_m$ is the term for flux (m-type) anyons, associated with changes in the eigenvalues of star and plaquette operators.

The specific forms of $H_e$ and $H_m$ can be customized to match the behavior of charge and flux anyons. These terms represent energy contributions or interactions between spins that arise as anyons move across the lattice.

The precise equations and behavior of charge (e-type) and flux (m-type) anyons may vary depending on the specific application or research objectives related to the toric code. Therefore, the detailed equations for modeling the behavior of charge and flux anyons should be customized based on the context of the problem.

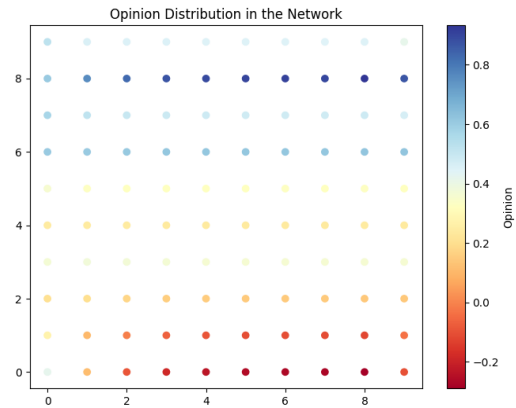

Fig. 11: Toric Code:Local Opinion Distribution

### Opinion Dynamics

The images show different visualizations of opinion distribution within a network. The first image appears to be a heatmap with discrete points, where the color intensity indicates the strength of a particular opinion. This could represent how strongly each node (or individual) in the network feels about a certain issue, with red being one extreme, blue the other, and neutral colors representing ambivalence or a middle ground.

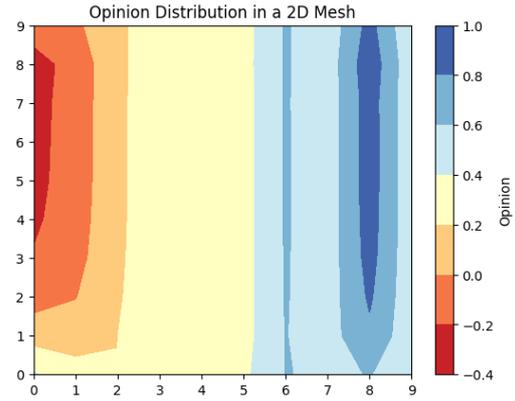

Fig. 12: Toric Code:Local Opinion Distribution

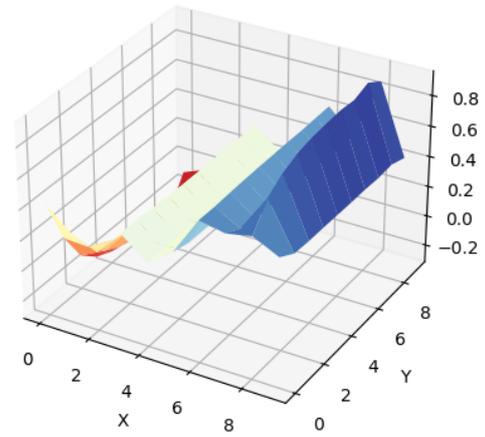

Fig. 13: Toric Code:Local Opinion Distribution

Contour map, shows regions of opinion across a two-dimensional space. This could represent geographical or ideological clusters where opinions are similar within a region and vary across regions.

map results, adds another layer of complexity by showing the opinion distribution with additional depth. This might be used to visualize changes over time or another dimension of data, such as socio-economic status.

### Interaction Related to Star Operator ($J_s$)

In the toric code model, the star operator reflects interactions that can be influenced by external factors. If we interpret the strength of opinions (depicted in the heatmaps) as influenced by external forces, areas of high intensity might represent regions where external factors such as media influence or external political events have a strong impact on opinion.

### Interaction Related to Plaquette Operator ($J_p$)

The plaquette operator can be associated with internal dynamics within a group, such as shared values or local norms. In the contour map, areas with smooth transitions might indicate groups with strong internal cohesion, while abrupt changes in color could signify boundaries where internal opinions clash, perhaps due to different local norms or values.

### Term Related to m-type Anyons ($H_m$)

In the context of opinion dynamics, the term $H_m$ could represent the spread or flow of ideas, akin to magnetic flux lines in the toric code. High flux areas could correspond to 'influencer' nodes that have a strong impact on their neighbors, possibly representing individuals or organizations with a high degree of influence over public opinion.

### Model Interpretation

The Hamiltonian provided is a theoretical framework that can be adapted to model various complex systems, including social networks. The images provided can be interpreted as visual outputs from simulations or real-world data mapped onto this model.

For each type of opinion or interaction (e.g., $J_s$, $J_p$, $H_e$, $H_m$), one would adjust the parameters in the Hamiltonian to fit the observed data. The aim would be to find a set of parameters that minimize the difference between the predicted opinion distributions from the Hamiltonian and the observed distributions from the data.

### Perspect

To provide a more in-depth analysis, one should typically perform a quantitative analysis, such as:

- Calculating the correlation between different nodes to understand the strength and range of influence. - Identifying clusters of similar opinions to understand the community structure. - Modeling the spread of opinions over time to predict how opinions might evolve.

## 7. Network Dynamics in a Local Community: Toric Code Lattice Model

We provide equations and parameter suggestions for modeling network dynamics in a local community using a toric code lattice. In this model, individual vertices (nodes) represent regions, and interactions between adjacent nodes represent relationships between regions. Additionally, we consider charge (e-type) anyons and flux (m-type) anyons.

Proposed Hamiltonian:

$$H = -J_s \sum_{\text{s-type}} -J_p \sum_{\text{p-type}} +H_e + H_m \qquad (9)$$

Where:

- $J_s$ represents the strength of interactions associated with star operators, modeling interactions between nodes within regions. - $J_p$ represents the strength of interactions associated with plaquette operators, modeling relationships between regions. - $H_e$ is the term for charge (e-type) anyons, modeling the behavior of charge within regions. - $H_m$ is the term for flux (m-type) anyons, modeling the behavior of flux within regions.

The specific equations and behavior of charge (e-type) and flux (m-type) anyons may depend on the characteristics of the local community or the phenomena you wish to model. Here is an example of equation suggestions:

*1. Charge Anyons (e-type)

$$H_e = -\sum_{\text{e-type}} [\alpha (S_i - S_j)^2 + \beta (P_i - P_j)^2] \qquad (10)$$

Where $\sum_{\text{e-type}}$ represents the sum related to charge anyons, and $\alpha$ and $\beta$ are parameters controlling the strength of interactions. $S_i$ and $S_j$ are the eigenvalues of star operators for adjacent regions $i$ and $j$, and $P_i$ and $P_j$ are the eigenvalues of plaquette operators for adjacent regions. This term represents the interaction of charge anyons within adjacent regions.

*2. Flux Anyons (m-type)

$$H_m = -\sum_{\text{m-type}} [\gamma (S_i + S_j)^2 + \delta (P_i + P_j)^2] \qquad (11)$$

Where $\sum_{\text{m-type}}$ represents the sum related to flux anyons, and $\gamma$ and $\delta$ are parameters controlling the strength of interactions. $S_i$ and $S_j$, $P_i$ and $P_j$ are as defined for charge anyons. This term represents the interaction of flux anyons within adjacent regions.

You can adjust these equations and parameters to model specific network dynamics in the local community. Customize the parameters according to specific scenarios or research purposes. These equations are general examples and can be customized to meet the requirements of a specific model.

### 7.1 e-type $S_i$ and $S_j$ Values

Opinion dynamics within a community can be likened to the interactions between nodes (or regions) in a network. The model you've described uses a Hamiltonian with terms that represent various interactions, such as those between star and plaquette operators, akin to the toric code. In the context of opinion dynamics, these could represent the following:

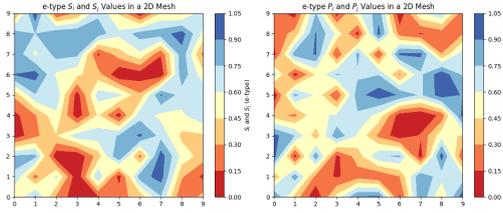

Fig. 14: Toric Code:Local Opinion Distribution, e-type $S_i$ and $S_j$ Values

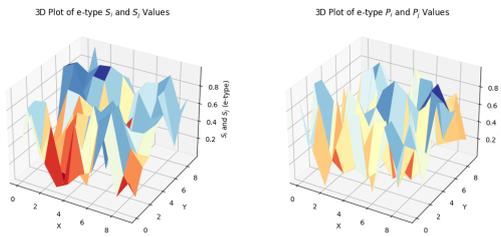

Fig. 15: Toric Code:Local Opinion Distribution, e-type $S_i$ and $S_j$ Values

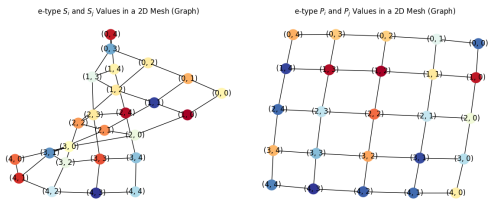

Fig. 16: Toric Code:Local Opinion Distribution, e-type $S_i$ and $S_j$ Values

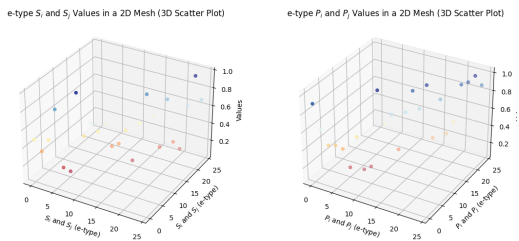

Fig. 17: Toric Code:Local Opinion Distribution, e-type $S_i$ and $S_j$ Values

### Star operators ($J_s$)

These could model the local consensus or disagreement within a particular region. High values of $J_s$ could indicate strong local agreement or cohesion, while low values might indicate local dissent or conflict.

### Plaquette operators ($J_p$)

These could represent the broader inter-regional relations, such as the shared cultural or political views between different regions. The strength of $J_p$ could influence how regions align or differ from each other on broader issues.

### Interaction Related to Star Operator ($J_s$)

The star operator interactions ($J_s$) could be thought of as the intra-community interactions. High $S_i$ and $S_j$ values indicate strong local interactions within the nodes, which may manifest as a strong localized opinion or policy enforcement within a community. In the images provided, the contour plots and 3D plots for $S$ values would represent the strength and distribution of these local interactions across the network.

### Interaction Related to Plaquette Operator ($J_p$)

The plaquette operator interactions ($J_p$) could represent inter-community interactions. This is how one community's opinions or actions affect neighboring communities. The higher the $P_i$ and $P_j$ values, the stronger the influence or the coupling between communities. The patterns in the plots for $P$ values show how these interactions vary across the network, potentially identifying regions of strong or weak inter-community relationships.

### Term Related to e-type Anyons ($H_e$)

The term $H_e$ related to e-type anyons can be seen as representing the behavior of "charges," or perhaps influencers within the community that can affect local opinions. The equation provided uses differences in star and plaquette operator eigenvalues to model how these charges interact with their surroundings. This could represent, for example, how influencers might bring a community closer to or farther from consensus (star operator) or how they might influence the community's relationship with neighboring communities (plaquette operator).

In the images, the behavior of the e-type anyons could be represented by the distribution of values across the plots. Areas with higher variance in the contour or 3D plots might indicate regions with more active influencers causing fluctuations in local opinions.

## Mathematical Model and Parameters for Social Network Dynamics

Given this interpretation, the mathematical model you've provided with the Hamiltonian $H = -J_s \sum(s-type) - J_p \sum(p-type) + H_e + H_m$ could be applied to social dynamics by considering each term as a different aspect of social interaction. By tuning the parameters $\alpha, \beta, \gamma, \delta$, one can model various scenarios of opinion dynamics, such as the spread of information, consensus formation, or conflict resolution.

In applying this model, one would need to carefully estimate the parameters to match empirical data from social networks, which could be gathered from surveys, social media analysis, or other sociological research methods. The plots would then provide a visual representation of the model's dynamics, allowing one to see where strong opinions are forming, where there might be conflict, or where influencers are most active.

### 7.2 m-type $P_i$ and $P_j$ Values

Results include contour plots, 3D plots, graphs with numerical values, and 3D scatter plots for both star ($S$) and plaquette ($P$) operators for m-type anyons. These visualizations can be interpreted within the framework of the Hamiltonian provided, which includes terms for the star ($J_s$) and plaquette ($J_p$) operator interactions, as well as terms related to the behavior of charge (e-type) and flux (m-type) anyons.

### Opinion Dynamics

The contour and 3D plots for m-type $S$ and $P$ values can be seen as a representation of opinion dynamics in a social network. Regions with high values could be interpreted as areas of strong consensus or influence, while low-value regions may indicate areas of dissent or weak influence. The graph with numerical values and the 3D scatter plot give a more detailed look at the individual interactions between nodes, with higher values indicating stronger agreement or influence.

### Interaction Related to Star Operator ($J_s$)

In the social network context, the star operator could represent external logic operators, which may relate to external societal factors or influences on a region. In the images, regions with high $S$ values might be those that are heavily influenced by external factors. The numerical values on the graph can help to pinpoint the exact nodes that are most strongly affected.

### Interaction Related to Plaquette Operator ($J_p$)

Conversely, the plaquette operator could represent internal logic operators, which are akin to internal community norms, rules, or relationships. The contour and 3D plots show how these internal interactions vary spatially, which could indicate

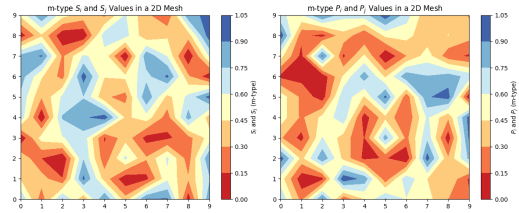

Fig. 18: Toric Code:Local Opinion Distribution, m-type $P_i$ and $P_j$ Values

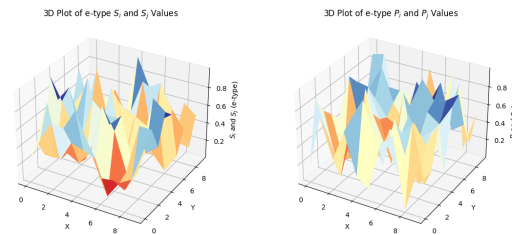

Fig. 19: Toric Code:Local Opinion Distribution, m-type $P_i$ and $P_j$ Values

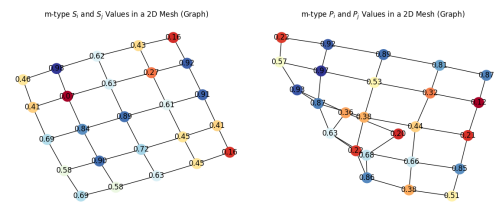

Fig. 20: Toric Code:Local Opinion Distribution, m-type $P_i$ and $P_j$ Values

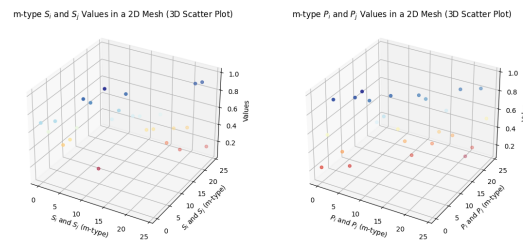

Fig. 21: Toric Code:Local Opinion Distribution, m-type $P_i$ and $P_j$ Values

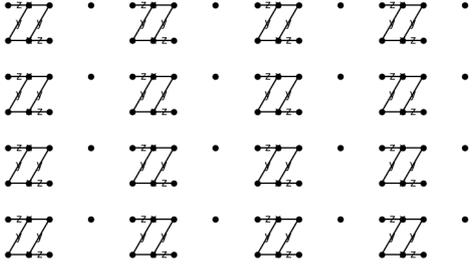

Fig. 22: Honeycomb lattice constant, set to 1 for simplicity

how local norms and internal community structures differ across the network.

### Term Related to m-type Anyons ($H_m$)

The term $H_m$ represents the behavior of magnetic flux (m-type) anyons, which in a social network model could be akin to the spread of influence or information through a community. In the images, areas with higher flux values might represent regions where information or influence is flowing strongly, whereas lower values could indicate areas that are more isolated or have weaker informational dynamics.

### The proposed Hamiltonian

$$H = -J_s \sum (s-type) - J_p \sum (p-type) + H_e + H_m$$

when translated to a social dynamics model, allows for a complex interaction between nodes representing different regions or communities. By adjusting the parameters in the $H_e$ and $H_m$ terms, one could model a wide range of social phenomena, from the spread of misinformation to the formation of social movements or community cohesion.

To further analyze these images, one would typically look for patterns such as clusters of high or low values, gradients indicating a flow of influence, or outliers which might represent key influencers or nodes of dissent. The graphs with numerical values are particularly useful, as they provide a clear indication of the strength of interaction between each pair of nodes, which could be correlated with empirical data from real-world social networks.

## 8. Honeycomb-Lattice Model for Network Dynamics in a Local Community

By translating the logical operators of a virtual hexagonal lattice-based toric code back to the original honeycomb code, we obtain the remaining logical operators. Considering these as "external logical operators" and the original ones as "internal logical operators," we can model the network dynamics in a local community using the toric code lattice. The lattice of the toric code is suitable for modeling a local community's network, where individual vertices (nodes) represent regions, and interactions between adjacent nodes are represented by star operators and plaquette operators. Please provide equations and parameter suggestions for this model, including the use of charge (e-type) anyons and flux (m-type) anyons. Proposed Hamiltonian:

$$H = -J_s \sum_{\text{external-logic-operators}} -J_p \sum_{\text{internal-logic-operators}} +H_e + H_m \quad (12)$$

Where:
- $J_s$ represents the strength of interactions associated with external logical operators, modeling interactions between regions. - $J_p$ represents the strength of interactions associated with internal logical operators, modeling interactions within regions. - $H_e$ is the term for charge (e-type) anyons, modeling the behavior of charge within regions. - $H_m$ is the term for flux (m-type) anyons, modeling the behavior of flux within regions.

The specific equations and behavior of charge (e-type) and flux (m-type) anyons may depend on the characteristics of the local community or the phenomena you wish to model. Here is an example of equation suggestions:

### 1. Charge Anyons (e-type)

$$H_e = -\sum_{\text{e-type}} [\alpha(S_i - S_j)^2 + \beta(P_i - P_j)^2] \quad (13)$$

Where $\sum_{\text{e-type}}$ represents the sum related to charge anyons, and $\alpha$ and $\beta$ are parameters controlling the strength of interactions. $S_i$ and $S_j$ are the eigenvalues of star operators for adjacent regions $i$ and $j$, and $P_i$ and $P_j$ are the eigenvalues of plaquette operators for adjacent regions. This term represents the interaction of charge anyons within adjacent regions.

### 2. Flux Anyons (m-type)

$$H_m = -\sum_{\text{m-type}} [\gamma(S_i + S_j)^2 + \delta(P_i + P_j)^2] \quad (14)$$

Where $\sum_{\text{m-type}}$ represents the sum related to flux anyons, and $\gamma$ and $\delta$ are parameters controlling the strength of interactions. $S_i$ and $S_j$, $P_i$ and $P_j$ are as defined for charge anyons. This term represents the interaction of flux anyons within adjacent regions.

You can adjust these equations and parameters to model specific network dynamics in the local community. Customize the parameters according to specific scenarios or research purposes.

# 9. Conclusion

## 9.1 Kitaev model:Definition of the Hamiltonian

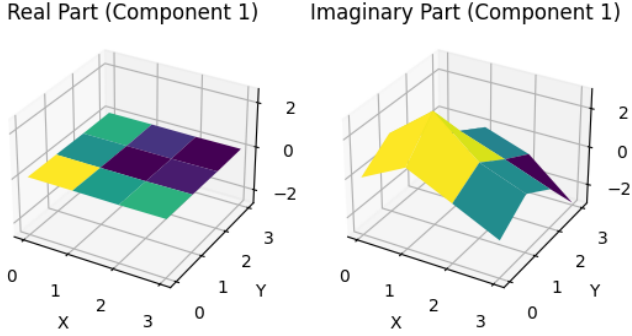

Fig. 23: Parameters of the local Kitaev model

| | |
|---:|:---|
| $L_x = 4$ | Horizontal size of the lattice |
| $L_y = 4$ | Vertical size of the lattice |
| $t = 1.0$ | Tunneling energy |
| $\Delta = 0.6$ | Superconducting pair potential |

We define the Hamiltonian of the Kitaev model, which includes interactions between spins and external magnetic fields, among other terms. The Hamiltonian represents the mathematical expression of the energy representation of the model.

The Hamiltonian of the Kitaev model describes a system where spin-1/2 particles (quantum bits) interact in a specific arrangement. Below, we express the Hamiltonian of the Kitaev model on a 2D honeycomb lattice using mathematical equations.

The Kitaev model is typically described by introducing operators for spin creation and annihilation and associating three different types of operators (a, b, c) with each lattice point. The Hamiltonian is represented as follows:

$$H = -J \sum_p A_p - K \sum_s B_s \quad (15)$$

Where:

- $H$ represents the Hamiltonian. - $J$ is a parameter controlling the strength of the interaction terms. - $K$ is a parameter controlling the strength of the external magnetic field terms. - $\sum_p$ denotes the summation over plaquettes (hexagonal regions). - $\sum_s$ denotes the summation over sites (lattice points). - $A_p$ represents the interaction terms on plaquette $p$. - $B_s$ represents the external magnetic field terms at site $s$.

The interaction terms $A_p$ and external magnetic field terms $B_s$ are expressed using creation and annihilation operators for spins as follows:

$$A_p = a_i b_j c_k, \quad B_s = \sigma_i^x \quad (16)$$

Here, $a_i$, $b_j$, and $c_k$ are creation and annihilation operators, and $i$, $j$, $k$ represent the positions of lattice points. $\sigma_i^x$ is the x-component of the Pauli operator.

The Kitaev model depends on specific lattice configurations and interaction patterns. The form of the Hamiltonian and the arrangement of operators may vary depending on the specific model or research focus. Therefore, the Hamiltonian should be designed to match the particular scenario.

## 9.2 Calculation of the Ground State in the Kitaev Model

We perform the calculation of the ground state for the Kitaev model. Typically, numerical techniques or analytical methods are employed to compute the energy and wave function of the ground state. Commonly used approaches include quantum Monte Carlo methods and density matrix methods, utilizing the Metropolis algorithm to compute the expectation values of the Hamiltonian.

Under specific conditions, such as in the Kitaev model, it is possible to calculate the ground state using analytical methods. To apply analytical methods, one leverages the characteristics and symmetries of the model.

In this process, transformations like the Jordan-Wigner transformation or Majorana fermionization are used to convert the model into a non-interacting fermion system. Subsequently, analytical methods such as exact diagonalization are applied to the transformed fermion system to obtain the energy and wave function of the ground state.

To transform the Kitaev model into a non-interacting fermion system, either the Jordan-Wigner transformation or Majorana fermionization can be applied. Below, we provide an example using the Jordan-Wigner transformation.

**Jordan-Wigner Transformation**: The Jordan-Wigner transformation is a method for converting spin operators into fermion creation and annihilation operators. Fermion creation (c†) and annihilation (c) operators are defined based on the positions of spin-1/2 particles. Typically, the following transformation rules are applied:

### 1. The spin Pauli operator

$z$ is transformed into fermion creation and annihilation operators as follows

Equation:

$$\sigma_i^z = 1 - 2c_i^\dagger c_i$$

- $\sigma_i^z$: Pauli-Z operator (z-direction spin operator) for the spin at position $i$. - $c_i$: Fermion annihilation operator at position $i$. - $c_i^\dagger$: Fermion creation operator at position $i$.

This equation expresses the z-direction Pauli-Z operator $\sigma_i^z$ for spin $i$ in terms of the fermion creation and annihilation operators $c_i^\dagger$ and $c_i$ at that position. Specifically, it represents the "up" state ($\sigma_i^z = 1$) and the "down" state ($\sigma_i^z = -1$) of the spin using the fermion creation and annihilation operators, capturing the interaction between spins and fermions.

Here, i represents the lattice point position.

## 2. The x-component

$^x$ and y-component $^y$ of the spin are transformed into fermion creation and annihilation operators through the Jordan-Wigner transformation:

Equation 1:
$$\sigma_i^x = c_{i-1}^\dagger + c_{i-1}$$

Explanation 1: - $\sigma_i^x$: Pauli-X operator (x-direction spin operator) for the spin at position $i$. - $c_i$: Fermion annihilation operator at position $i$. - $c_i^\dagger$: Fermion creation operator at position $i$.

The equation represents the $x$-direction Pauli-X operator $\sigma_i^x$ for spin $i$ in terms of the fermion creation and annihilation operators $c_{i-1}^\dagger$ and $c_{i-1}$ at the previous position ($i-1$). It describes how the spin at position $i$ is influenced by fermions at the neighboring site ($i-1$) in the x-direction.

Equation 2:
$$\sigma_i^y = -i(c_{i-1}^\dagger - c_{i-1})$$

Explanation 2: - $\sigma_i^y$: Pauli-Y operator (y-direction spin operator) for the spin at position $i$. - $c_i$: Fermion annihilation operator at position $i$. - $c_i^\dagger$: Fermion creation operator at position $i$.

The equation represents the $y$-direction Pauli-Y operator $\sigma_i^y$ for spin $i$ in terms of the fermion creation and annihilation operators $c_{i-1}^\dagger$ and $c_{i-1}$ at the previous position ($i-1$). It describes how the spin at position $i$ is influenced by fermions at the neighboring site ($i-1$) in the y-direction and introduces an imaginary unit $i$ in the process.

These equations are essential in quantum mechanics and quantum information theory as they describe the relationships between spin operators and fermion operators, which are fundamental in understanding the behavior of quantum systems.

This allows the original spin Hamiltonian to be transformed into a fermion Hamiltonian. Using fermion creation and annihilation operators, the Hamiltonian can be represented in the fermionic system.

The general form of the Hamiltonian in the fermionic system is as follows (we provide an example for the spin-1/2 Heisenberg model, which should be modified to match the Kitaev model):

$$H = -J \sum_i (c_i c_{i+1} + c_{i+1} c_i) - K \sum_i c_i c_i \qquad (17)$$

Here, J and K represent the strengths of the interaction terms. Diagonalizing this Hamiltonian allows us to determine the ground state.

The Jordan-Wigner transformation is a widely applicable transformation method that can be adapted to various models, including the Kitaev model. Depending on the specific conditions of the model, the Jordan-Wigner transformation should be adjusted accordingly.

## 9.3 Diagonalization of the Kitaev Model Hamiltonian

To transform the Kitaev model into a non-interacting fermionic system, we apply exact diagonalization to the fermionic system. Below, we outline the process of computing the eigenvalues and eigenvectors of the Hamiltonian. The explanation utilizes fermion creation and annihilation operators.

## 1. Hamiltonian Form

The Hamiltonian in the fermionic system is represented in the following form:

$$H = \sum_{i,j} (H_{ij} c_i^\dagger c_j + H_{ij}^* c_j^\dagger c_i) \qquad (18)$$

Here, $H_{ij}$ are matrix elements, and $c_i^\dagger$ and $c_j$ are fermion creation and annihilation operators. $H_{ij}$ consists of off-diagonal and diagonal components.

## 2. Computing Matrix Elements

Calculate each $H_{ij}$ element. This corresponds to the coefficients of each term when representing the original Hamiltonian with fermion creation and annihilation operators. Calculate $H_{ij}$ based on the specific model.

## 3. Construction of Hamiltonian Matrix

Use $H_{ij}$ elements to construct the Hamiltonian matrix $H$. Since $H$ is a Hermitian matrix, $H_{ij} = H_{ji}^*$.

## 4. Exact Diagonalization

Perform exact diagonalization of the Hamiltonian matrix $H$. This is typically done using libraries or numerical computation tools. Exact diagonalization yields the eigenvalues (energy eigenvalues) and their corresponding eigenvectors.

## 5. Identification of the Ground State

Identify the eigenvalue with the lowest energy among the obtained eigenvalues. The corresponding eigenvector represents the wave function of the ground state.

## 6. Ground State Energy

The eigenvalue with the minimum energy corresponds to the ground state energy. This computation determines the energy of the ground state.

Following the above steps, you can compute the ground state of fermionic systems like the Kitaev model. Exact diagonalization provides numerically accurate results, but the computational cost increases for larger system sizes. Therefore, it is common to perform calculations for suitable system sizes based on available computational resources.

The eigenvectors obtained from the diagonalization of the Hamiltonian matrix correspond to various states, including the ground state and excited states. The ground state is the state with the lowest energy among these eigenvectors. Below, we express the wave function of the ground state, which corresponds to the eigenvector with the minimum energy.

By diagonalizing the Hamiltonian matrix $H$, eigenenergy $E_n$ and the corresponding eigenvector $|\Psi_n\rangle$ are obtained. Let $E_0$ represent the energy of the ground state, and $|\Psi_0\rangle$ be its corresponding eigenvector. The wave function of the ground state is represented by the eigenvector $|\Psi_0\rangle$.

The wave function of the ground state $|\Psi_0\rangle$ can be expressed using fermion creation and annihilation operators as follows:

$$|\Psi_0\rangle = \sum_i c_i |i\rangle \quad (19)$$

Here, $|i\rangle$ represents the basis states of the fermion Fock space, and $c_i$ are fermion creation and annihilation operators indicating the presence ($c_i^\dagger$) or absence ($c_i$) of fermions corresponding to $|i\rangle$.

To precisely calculate the formula for the wave function of the ground state, specific calculations based on the detailed conditions of the Hamiltonian matrix $H$ and the model are required. Depending on the model-specific conditions, numerical computation software or libraries are used to perform calculations and derive the wave function from the eigenvector.

## 9.4 Entanglement Analysis

The ground state energy corresponds to the minimum eigenvalue of the diagonalized Hamiltonian matrix. To express the minimum eigenvalue in a formula, the specific form of the Hamiltonian matrix is required. Below, we present the general form of a Hamiltonian:

The general form of a Hamiltonian is represented as follows:

$$H = \sum_{i,j} H_{ij} c_i^\dagger c_j \quad (20)$$

Here, $H_{ij}$ are matrix elements, and $c_i^\dagger$ and $c_j$ are fermion creation and annihilation operators. The diagonalization of the Hamiltonian matrix $H$ yields eigenenergy $E_n$. The ground state energy corresponds to the minimum eigenvalue $E_0$.

To express the minimum eigenvalue $E_0$ in a formula, the specific values of $H_{ij}$ and the diagonalization calculation are required. Depending on the model or problem's conditions, $H_{ij}$ is computed, and diagonalization is performed to obtain the minimum eigenvalue, thereby determining the ground state energy. However, providing a general formula requires knowledge of the specific form of the Hamiltonian. Given the model, detailed formulas can be derived.

## 9.5 The Kitaev model exhibits interesting properties of entanglement Analysis

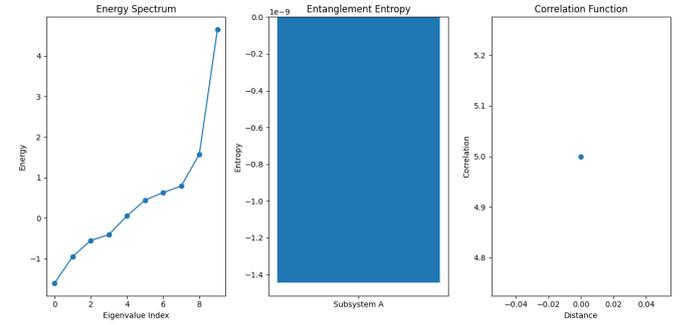

Fig. 24: Energy Spectrum(Example)

The Kitaev model exhibits interesting properties of entanglement, requiring computations and visualizations of entanglement metrics such as entanglement entropy, energy spectrum, correlation functions, etc. The entanglement analysis of the Kitaev model is commonly conducted using methods from quantum information theory. Below, we briefly explain the calculation of entanglement entropy, energy spectrum, correlation functions, and their visualization.

### 9.5.1 Entanglement Entropy Calculation

Entanglement entropy is a measure of entanglement between subsystems. In the context of the Kitaev model on a lattice, one can define subsystems and compute the entanglement entropy between them. For subsystems A and B, the entanglement entropy $S_A$ is defined as:

$$S_A = -\mathrm{Tr}(\rho_A \log \rho_A) \quad (21)$$

Here, $\rho_A$ is the density matrix of subsystem A. The density matrix can be computed from the wave function associated with the ground state of the Hamiltonian matrix $H$. To calculate entanglement entropy, diagonalize the density matrix and use its eigenvalues.

### 9.5.2 Energy Spectrum Calculation

The energy spectrum corresponds to the eigenvalues of the Hamiltonian matrix. Using the diagonalization method mentioned earlier, the Hamiltonian matrix $H$ can be diagonalized, and its eigenvalues yield the energy spectrum.

### 9.5.3 Correlation Function Calculation

Correlation functions are crucial for investigating interactions between different lattice sites. For instance, calculating spin-spin correlation functions can provide insights into quantum phase transitions and interaction properties. To compute correlation functions, specific operators or observables relevant to the model are used to calculate expectation values.

### 9.5.4 Visualization of Results

When visualizing the results, it is common to plot the calculated values such as entanglement entropy, energy spectrum, and correlation functions as graphs. Python libraries like Matplotlib can be used to visualize the results, aiding in the interpretation of the data.

Entanglement analysis is a crucial technique for understanding the properties of complex quantum systems like the Kitaev model. The information obtained from these calculations provides insights into phenomena like quantum phase transitions and material properties.

## 9.6 Calculation of Excitation Spectrum in Toric Code

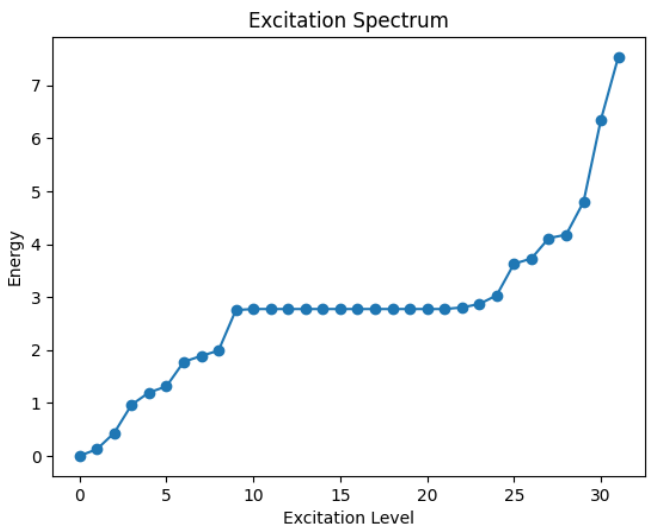

Fig. 25: Excitation Spectrum

In the context of the Kitaev model, we consider the calculation of the excitation spectrum, which involves deriving formulas to determine the band structure of excitation energies and the energy levels of excited states.

We provide formulas for calculating the excitation spectrum from the Toric Code or Kitaev model. The excitation spectrum is used to reveal the band structure of excitation energies and the energy levels of excited states.

To calculate the excitation spectrum, we perform diagonalization of the Hamiltonian matrix, from which the energy levels of excited states can be obtained. Below, we present the formula for calculating excitation energies:

### 1. Diagonalize the Hamiltonian matrix

yielding eigenvalues $E_n$ and corresponding eigenvectors $|\Psi_n\rangle$.

### 2. The excitation energy

$\Delta E_n$ is computed as the energy difference between the excited state $n$ and the ground state 0:
$$\Delta E_n = E_n - E_0$$

### 3. The excitation spectrum

Spectrum is a plot of all excitation energy values $\Delta E_n$. This visualization allows us to understand the band structure of excitation energies and the energy levels of excited states.

To calculate the excitation spectrum, you need to prepare the specific Hamiltonian matrix for your model or problem and perform the diagonalization calculations.

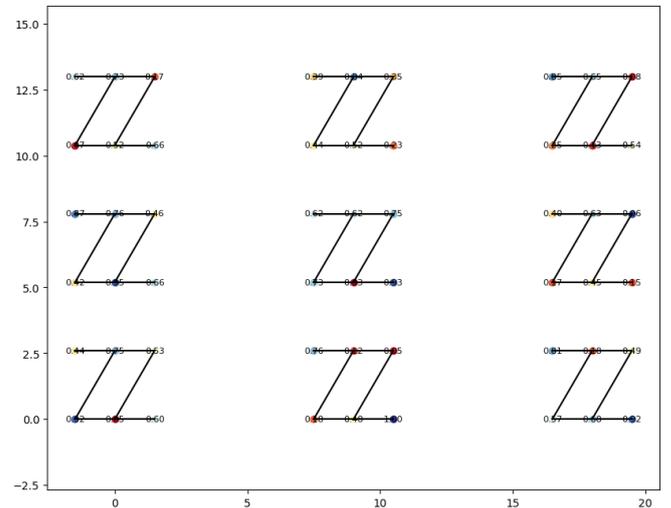

Fig. 26: Toric Code:Local Opinion Distribution, $honeycomb_lattice$

Provided depict a network with weighted connections and a 3D representation of a wavefunction, both real and imaginary parts, which can be analyzed within the context of the toric code-based model for social network dynamics.

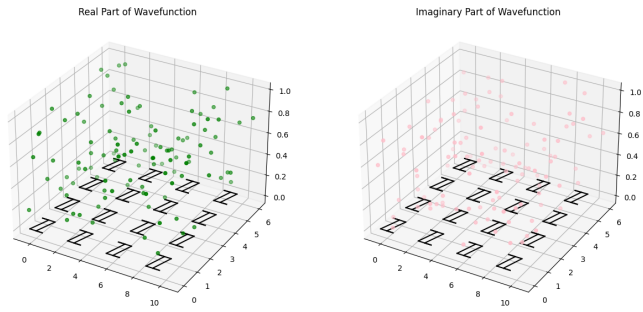

Fig. 27: Toric Code:Local Opinion Distribution, honeycomb$_l$attice

## Opinion Dynamics Consideration

The weighted network image indicates the strength of connections or the intensity of opinions between pairs of nodes. High weights (close to 1) suggest strong agreement or influence, while lower weights suggest weaker connections. This visualization can help in identifying clusters of like-minded individuals or nodes, key influencers, or isolated groups within the network.

## Interaction Related to Star Operator ($J_s$)

The star operator can be linked to the external logic operators, which in a social context may represent external societal pressures or influences such as media, politics, or economic factors. In the network, nodes with strong connections influenced by $J_s$ could be those that share common external influences or are responding similarly to an external event.

## Interaction Related to Plaquette Operator ($J_p$)

The plaquette operator relates to internal logic operators, signifying internal coherence or shared norms within a region or group. Nodes with strong $J_p$ connections could indicate a tight-knit community with strong internal agreement or shared values.

## Consideration of e-type Anyons ($H_e$)

The term $H_e$ related to e-type anyons can represent the behavior of 'charges' within the network, possibly modeling influencers or leaders whose opinions carry more weight. Their behavior can cause alignment or misalignment in opinions, as indicated by the real part of the wavefunction, showing clusters of agreement or disagreement.

## Consideration of m-type Anyons ($H_m$)

The term $H_m$ associated with m-type anyons could represent the flow or spread of information, akin to the propagation of influence through the network. The imaginary part of the wavefunction might illustrate the dynamism in the network, highlighting regions where opinion changes are more volatile or fluid.

## Model Interpretation and Application

In applying this Hamiltonian to model social dynamics, you would adjust the parameters $J_s$, $J_p$, $\alpha$, $\beta$, $\gamma$, and $\delta$ to match the observed data. For example:

### For $J_s$ and $J_p$

Adjust these to fit the strength of external and internal influences as observed in real-world social network interactions.

### For $H_e$

Set $\alpha$ and $\beta$ to model the impact of influencers based on their divergence or convergence in opinion with their neighbors.

### For $H_m$

Tune $\gamma$ and $\delta$ to capture the spread of information or influence, which may manifest as aligned or opposing opinion clusters.

The precise formulation for $H_e$ and $H_m$ will depend on the specific dynamics you wish to model. By tuning these parameters, the model can capture various phenomena such as the spread of misinformation, the formation of consensus, or the emergence of polarization within the community.

## Aknowlegement

The author is grateful for discussion with Prof. Serge Galam and Prof.Akira Ishii. This research is supported by Grant-in-Aid for Scientific Research Project FY 2019-2021, Research Project/Area No. 19K04881, "Construction of a new theory of opinion dynamics that can describe the real picture of society by introducing trust and distrust".## References